\documentclass{iopart}

\usepackage{graphicx,epstopdf}
\usepackage{psfrag}
\usepackage{enumerate}

\renewcommand{\u}{\underline}
\renewcommand{\b}{\mathbf}
\renewcommand{\o}{\overline}
\renewcommand{\hat}{\widehat}
\renewcommand{\t}{\widetilde}
\renewcommand{\i}{\includegraphics}

\begin{document}

\title{Dynamical nonlocal coherent-potential approximation for itinerant electron magnetism}

\author{D~A~Rowlands and Yu-Zhong~Zhang$^{*}$}
\address{Shanghai Key Laboratory of Special Artificial Microstructure Materials and Technology,
School of Physics Science and Engineering, Tongji University, Shanghai 200092, P.R.~China}
\ead{$^{*}$yzzhang@tongji.edu.cn}

\begin{abstract}
A dynamical generalisation of the nonlocal coherent-potential approximation is derived based upon the functional integral approach to the interacting electron problem. The free energy is proven to be variational with respect to the self-energy provided a self-consistency condition on a cluster of sites is satisfied. In the present work, calculations are performed within the static approximation and the effect of the nonlocal physics on the formation of the local moment state in a simple model is investigated. The results reveal the importance of the dynamical correlations.
\end{abstract}


\submitto{\JPCM}
%--------------------------------------------------------------------------------------------------------------------------------------------
\section{Introduction}

Until the introduction of the single-site spin fluctuation (SSF) theory by Hubbard~\cite{Hubbard1979a,Hubbard1979b} and Hasegawa~\cite{Hasegawa1979,Hasegawa1980a,Hasegawa1980b} in the late 1970s, the development of a theory which could provide a satisfactory description of the magnetism of ferromagnetic metals had remained a long-standing problem~\cite{Moriya1985,Kakehashi2012}. In particular, the SSF was able to explain both the itinerant and localized aspects of the experimental data, for example the non-integer values of the magnetic moments normally associated with an itinerant model, and the Curie-Weiss behaviour of the magnetic susceptibility normally associated with a localized model~\cite{Moriya1985,Kakehashi2012}.

The SSF is based upon the functional integral method in which the Hubbard-Stratonovich transformation~\cite{Hubbard1959} is used to convert the interacting electron system into a non-interacting one-electron system in the presence of auxiliary charge and exchange fields. These fields vary randomly from site to site and are in principle functions of imaginary time $\tau$, respresenting the effect of the many-body interactions. Minimisation of the free energy will determine the behaviour of the system and thus the central quantity is the energy functional $E[\xi,\eta]$ for configurations of the random charge and exchange fields $\{\xi_i(\tau),\eta_i(\tau)\}$, where $i$ runs over all lattice sites and $\tau$ from $0\rightarrow\beta$. The energy functional is in principle needed to take a thermal average over all possible field configurations in terms of functional integrals. In the SSF, two major approximations are made. The first is to neglect the time-dependence of the fields so that they become static quantities. The second is to introduce an effective medium and use the coherent-potential approximation (CPA)~\cite{Hubbard1964,Soven1967} to perform the averaging over the field configurations. Since the CPA is a local approximation, only field configurations on one impurity site need to be considered and the medium can be determined self-consistently. The overall physical picture is that of relatively fast moving itinerant electrons which travel from site to site causing fluctuations in the total atomic spins. These spin fluctuations self-consistently maintain the auxiliary exchange fields which vary relatively slowly in magnitude and direction and act as effective ``local moments''~\cite{Hubbard1979a,Hubbard1979b}. Indeed, the idea of the SSF was also used to develop the first-principles ``disordered local moment'' (DLM) theory of finite-temperature metallic magnetism based on one-electron potentials provided by density functional theory and implemented within the KKR-CPA band-structure method~\cite{Gyorffy1985,Staunton1992,Staunton2014}.

The successes of the SSF notwithstanding, quantitatively the results obtained were some way from the experimental data, for example calculated Curie temperatures were much too high. This was a consequence of both of the two major approximations made, i.e.~static and local. A major breakthrough was made in 1992 when Kakehashi~\cite{Kakehashi1992} lifted the limitations imposed by the static approximation by deriving the CPA equations within the functional integral scheme whilst maintaining the time-dependence of the auxiliary fields. The quantum Monte Carlo (QMC) method was employed to solve the dynamical impurity problem. The resulting dynamical CPA (dyn-CPA) method revealed that inclusion of dynamical correlations reduced the Curie temperatures by a factor of 2 as compared with the results of the SSF~\cite{Kakehashi1992,Kakehashi2002a}. Nevertheless, as dyn-CPA calculations have become quantitatively more accurate through interfacing with first-principles methods~\cite{Kakehashi2008}, it has been revealed that calculated Curie temperatures for simple metals such as Fe and Co are still 1.8 times higher than the experimental values~\cite{Kakehashi2011a,Kakehashi2011b}. The difference is believed to be due to the neglect of the nonlocal physics.

Around the same time the dyn-CPA was developed came the introduction of the dynamical mean-field theory (DMFT)~\cite{Georges1992} within the context of strongly-correlated electron systems which clarified the physics of the metal-insulator transition~\cite{Georges1996}. Subsequently much work was done in developing nonlocal generalizations of DMFT through the introduction of cluster methods such as the dynamical cluster approximation (DCA)~\cite{Hettler1998}. In the DCA, nonlocal spatial correlations are systematically taken into account as the cluster size is increased whilst the translational invariance of the underlying lattice is preserved from the outset. The idea of the DCA was subsequently used to generalize the conventional CPA method within the context of disordered alloy systems where it is known as the nonlocal CPA~\cite{Jarrell2001,Moradian2002,Rowlands2003}. Indeed, the development of a generalization of the CPA which preserves the translational invariance of the underlying lattice essential for the description of disordered alloys had remained a long-standing problem~\cite{Gonis1992,Rowlands2009}.

On the other hand, no such developments have been made within the functional integral scheme. The purpose of the present paper is to take the first step in this direction by deriving a dynamical nonlocal CPA (dyn-NLCPA) theory. Since the time-dependence of the auxiliary fields is retained, this is in effect a generalization of the dyn-CPA equations to include the effects of nonlocal spatial correlations in the field configurations. An important feature of the theory is that since the translational invariance of the underlying lattice is preserved, quantities such as the ``local moments'' represented by the exchange fields can retain their single-site description, the difference being that they are not formed independently from the surrounding sites as in the single-site CPA. Although only in an approximate way once computational schemes are introduced, the dyn-CPA and DMFT have been shown to be equivalent in principle~\cite{Kakehashi2002b} and hence the dyn-NLCPA can be viewed as the analog of the DCA within the functional integral method. 

As a first step in understanding the implications of the new theory, calculations are presented here for a simple model within the static approximation to the dyn-NLCPA. The aim is to investigate the effects of nonlocal spatial correlations on the formation of the ``local moment'' state in the absence of dynamical correlations. It will be shown that the static approximation is not adequate if nonlocal spatial correlations are included.

This paper is organised as follows. In section~\ref{fim} the basic idea of the functional integral method is outlined. In section~\ref{method} the dyn-NLCPA is derived. After obtaining an effective cluster model by coarse-graining in k-space, the free energy for an impurity cluster Hamiltonian is obtained. By demanding that the self-energy minimises the corresponding free energy, a self-consistency condition is obtained which requires the thermal average of the impurity Green's function with respect to a cluster energy functional. Following this, it is shown how properties are calculated before describing a means of solving the quantum cluster impurity problem. In section~\ref{static}, the static approximation to the dyn-NLCPA is described. In section~\ref{results}, the effect of nonlocal correlations in the exchange-field configurations on the formation of the local moment state in the paramagnetic regime is investigated for a simple model within the static approximation. Finally conclusions are made in section~\ref{conclusions}.

%--------------------------------------------------------------------------------------------------------------------------------------------
\section{Formalism}\label{formalism}

%-------------------------------------------------------------------
\subsection{Functional integral method}\label{fim}

The functional integral method is based upon the Hubbard-Stratonovich transformation
\begin{equation}\label{hs}
	e^{A\hat{O}^2}=\sqrt{\frac{A}{\pi}}\int{d\xi}e^{-A\xi^2+2A\hat{O}\xi}
\end{equation}
which linearizes the exponential of a quadratic operator in terms of a functional integration over random fields. If we consider the Hubbard Hamiltonian
\begin{equation}\label{h}
	H=H_0+H_I
\end{equation}
with non-interacting part
\begin{equation}
	H_0=\sum_{i,j,\sigma}t_{ij}a^{\dag}_{i\sigma}a_{j\sigma}
\end{equation}
where $t$ is the hopping parameter, and interacting part
\begin{equation}
	H_I=\sum_{i,\sigma}(\epsilon_0-\mu)n_{i\sigma}+\sum_i{U}n_{i\uparrow}n_{i\downarrow}
\end{equation}
where $\epsilon_0$ is an atomic energy, $\mu$ is the chemical potential, and $U$ is the on-site interaction, then $H_I$ needs to be re-expressed in a suitable form for the purposes of applying the Hubbard-Statonovich transformation (\ref{hs}). There is some scope in the form of the interaction that can be chosen~\cite{Prange1981}. In the present work the two-field method adopted by Hasegawa~\cite{Hasegawa1980b} is used such that
\begin{equation}\label{interaction}
	n_{i\uparrow}n_{i\downarrow}=\frac{1}{4}(n_i^2-m_{iz}^2)
\end{equation}
where the magnetization $m_i$ has been quantized in the z-direction. This has the advantage of a well-known static limit at the ground state, namely the Hartree-Fock approximation. On the other hand, it breaks the rotational invariance of the magnetic moments. If rotational invariance is required, for example to describe non-collinear magnetism, then the four-field method adopted by Hubbard~\cite{Hubbard1979a,Hubbard1979b} can be used instead.

In the interaction representation, the partition function corresponding to the Hamitonian (\ref{h}) is defined by
\begin{equation}
	Z=e^{-\beta{\cal{F}}}=\Tr\left[e^{-\beta{H_0}}\,{\cal{T}}\exp\left(-\int_0^\beta{H_I}(\tau)d\tau\right)\right],
\end{equation}
where ${\cal{T}}$ is the time-ordering operator and $\Tr$ denotes a trace over lattice site, spin and time. Discretizing the time interval $[0,\beta]$ into $N$ mesh points and applying the Hubbard-Statonovich transformation at each time value $\tau_n$ of infinitesimal duration $\Delta\tau=\beta/N$ yields the following expression for the partition function,
\begin{eqnarray}\label{partition}
	Z=e^{-\beta{\cal{F}}}=&\left[\prod_{i=1}^N\int\delta\xi_i(\tau)\delta\eta_i(\tau)\right]Z^1[\xi,\eta]\nonumber\\
		                              &\times\exp\left[-\frac{1}{4}U\sum_{i=1}^N\int_0^\beta{d\tau}\left\{\eta^2_i(\tau)+\xi^2_i(\tau)\right\}\right],
\end{eqnarray}
where $Z^1$ is the partition function for a one-body system,
\begin{equation}
	Z^1[\xi,\eta]=\Tr\left[{\cal{T}}\exp\left(-\int_0^\beta{d\tau}\,H^{1}\left(\tau,\xi(\tau),\eta(\tau)\right)\right)\right],
\end{equation}
and the functional integrals are defined by
\begin{equation}\label{funcinttime}
	\int\delta\xi_i(\tau)\equiv\lim_{N\rightarrow\infty}\int\left[\prod_{n=1}^{N}\sqrt{\frac{\beta{U}}{4\pi{N}}}\,d\xi_i(\tau_n)\right].
\end{equation}
The one-body Hamiltonian $H^{1}\left(\tau,\xi(\tau),\eta(\tau)\right)$ is defined by
\begin{equation}
	 {\fl}H^{1}\left(\tau,\xi(\tau),\eta(\tau)\right)=\sum_{i,\sigma}v_{i\sigma}\left(\xi_i(\tau),\eta_i(\tau)\right)n_{i\sigma}(\tau)+\sum_{i,j,\sigma}t_{ij}a^{\dag}_{i\sigma}(\tau)a_{j\sigma}(\tau)
\end{equation}
in terms of a random one-body potential
\begin{equation}\label{potential}
	v_{i\sigma}\left(\xi_i(\tau),\eta_i(\tau)\right)=\epsilon_0-\mu+\frac{1}{2}U\left(i\eta_i(\tau)-\sigma\xi_i(\tau)\right).
\end{equation}
At each site $i$ and time value, the one-body potential depends upon the values of the charge and exchange fields $\eta_i(\tau)$ and $\xi_i(\tau)$, respectively. The free energy is obtained by re-arranging (\ref{partition}) in the form
\begin{equation}\label{free}
	{\cal{F}}=-\frac{1}{\beta}\ln\left[\prod_i\int\delta\xi_i\delta\eta_i\right]e^{{-\beta}E[\xi,\eta]}
\end{equation}
with energy functional $E[\xi,\eta]$ defined by
\begin{equation}\label{energyfunctional}
	E[\xi,\eta]=-\frac{1}{\beta}\ln{Z^1}[\xi,\eta]+\sum_i\frac{U}{4\beta}\int_0^{\beta}d\tau\left\{\eta_i^2(\tau)+\xi_i^2(\tau)\right\}.
\end{equation}
The energy functional is a measure of the probability of the field variable values occuring in practice and is therefore central to the functional integral method.

The one-body partition function can also be expressed in the more explicit form~\cite{Wang1969}
\begin{equation}\label{lnZ1}
	\ln{Z^1}=\ln{Z^0}+\Tr[\ln{g}]+\Tr[\ln{G}^{-1}],
\end{equation}
where $Z^0=\Tr[e^{-\beta{H}_0}]$ is the free-electron partition function, $H_0$ is the free-electron Hamiltonian, $g$ is the free-electron Green's function, and $G$ is the temperature Green's function for the one-body potentials defined by
\begin{equation}
	G_{ij\sigma}(\tau,\tau')=-\frac{\Tr\left[{\cal{T}}a_{i\sigma}(\tau)a^{\dag}_{j\sigma}(\tau')e^{-\int_0^{\beta}H^{1}(\tau'')d\tau''}\right]}{\Tr\left[e^{-\int_0^{\beta}H^{1}(\tau'')d\tau''}\right]},
\end{equation}
which is diagonal in the spin indices. In (\ref{lnZ1}), the trace is over lattice site, time and spin. If the one-body problem could be solved exactly, the exact temperature Green's function $\o{G}_{ij\sigma}(\tau,\tau')$ could also be written in the form of a Dyson equation,
\begin{eqnarray}\label{Gexact}
 {\fl} \o{G}_{ij\sigma}(\tau-\tau')=g_{ij\sigma}(\tau-\tau') \nonumber\\	
       +\int_0^{\beta}d{\tau_1}d{\tau_2}\,\sum_{k}g_{ik\sigma}(\tau-\tau_1)\Sigma_{k\sigma}(\tau_1-\tau_2)\o{G}_{lj\sigma}(\tau_2-\tau'),
\end{eqnarray}
where $\Sigma_{ij\sigma}(\tau-\tau')$ is the exact nonlocal dynamical self-energy. Since these quantities are translationally-invariant in both space ($i-j$) and time ($\tau-\tau'$), a momentum-frequency transform may be used to express the above equation in momentum and frequency space as follows,
\begin{equation}\label{Gkspace}
  \o{G}_{\sigma}(\b{k},i\omega_l)=g_{\sigma}(\b{k},i\omega_l)+g_{\sigma}(\b{k},i\omega_l)\Sigma_{\sigma}(\b{k},i\omega_l)\o{G}_{\sigma}(\b{k},i\omega_l),
\end{equation}
where $\b{k}$ is a vector in the Brillouin zone (BZ) of the lattice and $\omega_l$ are the Matsubara frequencies.

%-------------------------------------------------------------------
\subsection{Dynamical nonlocal CPA}\label{method}

The dyn-NLCPA can be derived either in the imaginary time or the frequency representation. Here the frequency representation is adopted since this is the most useful for practical calculations. 

%----------------
\subsubsection{Effective cluster model}\label{coarse}

\begin{figure}
 \begin{center}
{\renewcommand{\arraystretch}{4}
 \begin{tabular}{c}
 \scalebox{0.7}{\i{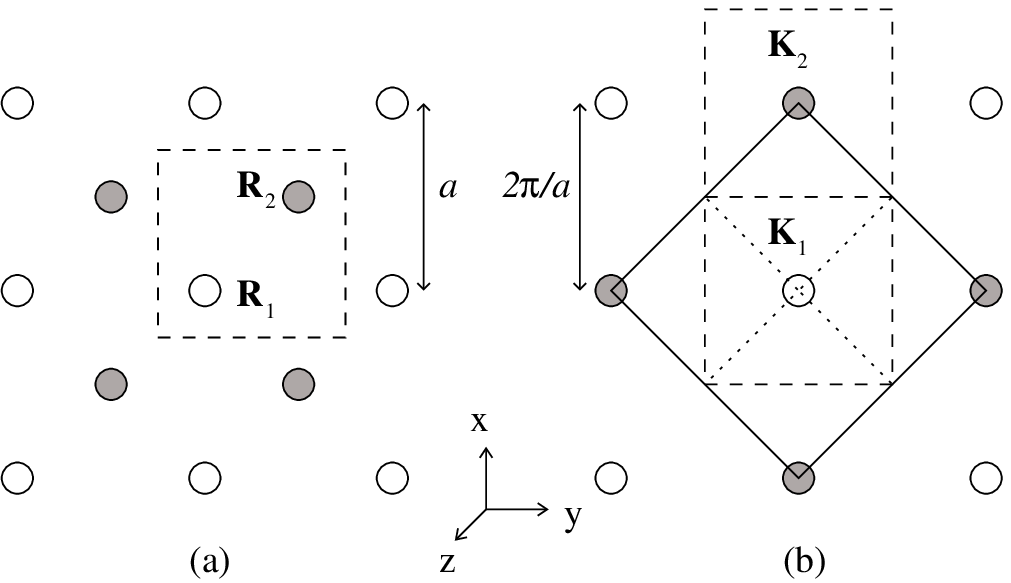}} \\ \scalebox{0.7}{\i{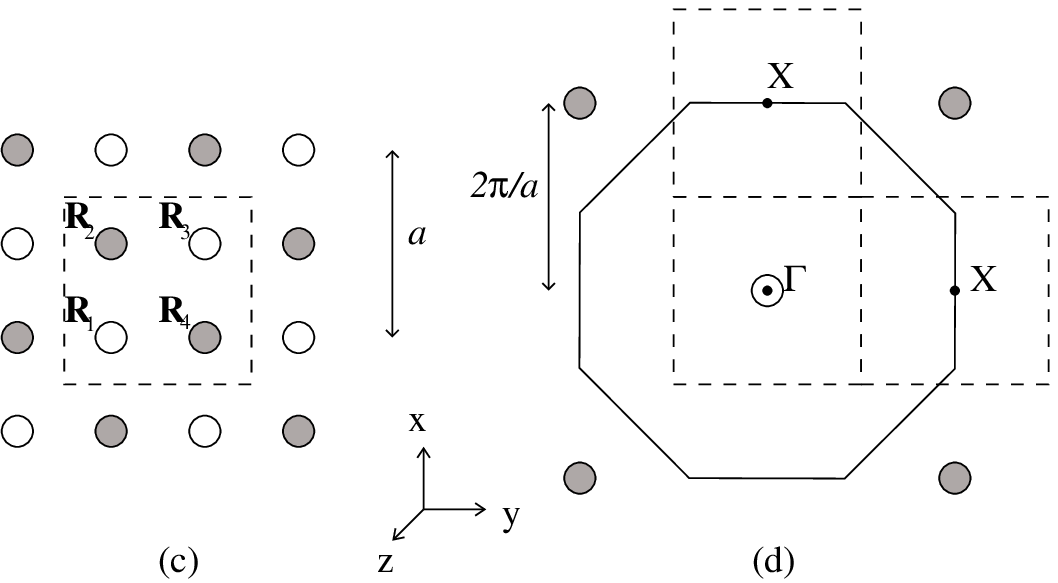}}
 \end{tabular}}
 \caption{\label{cells} (a) Cross-section of a real-space sublattice unit cell (dashed line) for the bcc lattice with $N_c=2$ cluster containing the points $\b{R}_1=(0,0,0)$ and $\b{R}_2=(a/2,a/2,a/2)$. The shaded sites lie out of the page. It may help to shift the cell so that it is centred at a sublattice site. (b) Cross-section of the corresponding reciprocal-space tiles (dashed lines) for the $N_c$=2 cluster, with $\b{K}_1=(0,0,0)$ and $\b{K}_2=(2\pi/a,0,0)$ at their centres. The shaded points lie out of the page and the solid line denotes a cross-section of the first BZ in the $(k_x,k_y)$ plane. The BZ can be visualised as a cube with a pyramid attached to each of the six faces, and the dotted line shows a projection of such a pyramid into the $k_z$ plane. (c) Cross-section of a real-space sublattice unit cell for a $N_c$=4 cluster on the fcc lattice containing the points $\b{R}_1=(0,0,0)$, $\b{R}_2=(a/2,0,a/2)$, $\b{R}_3=(a/2,a/2,0)$ and $\b{R}_4=(0,a/2,a/2)$. The shaded sites lie out of the page. Again it may help to shift the cell so that it is centred at a sublattice site. (d) Cross-section of the corresponding reciprocal-space tiles (dashed lines) for the $N_c$=4 cluster, with $\b{K}_1=(0,0,0)$, $\b{K}_2=(2\pi/a,0,0)$, and $\b{K}_3=(0,2\pi/a,0)$ shown as the $\Gamma$ point and the two $X$ points. The fourth tile is centered at the $X$ point $\b{K}_4=(0,0,2\pi/a)$ and is situated out of the page vertically above $\Gamma$. Again the shaded points lie out of the page and the solid line denotes a cross-section of the first BZ in the $(k_x,k_y)$ plane. Figure reproduced from reference~\cite{Rowlands2009}. }
 \end{center}
\end{figure}

The first step in the derivation is to consider the lattice in an equivalent representation as a set of $N_c$ sublattices~\cite{Rowlands2009}. Correspondingly the lattice BZ is divided into a set of sublattice BZs centred at the set of reciprocal sublattice vectors $\{\b{K}_n\}$. This procedure is illustrated in figure~\ref{cells} for the realistic bcc and fcc lattice structures needed for describing metals such as iron and nickel.

The first main approximation made is to approximate the exact lattice self-energy of (\ref{Gkspace}) (which in the present context is fully nonlocal and dynamical) as a constant in each sublattice BZ so that it becomes a step function in momentum space. The dynamical properties of the self-energy are retained so that it remains translationally-invariant in both space and time. A simplistic physical interpretation is that spatial correlations between sites on the same sublattice are neglected and only spatial correlations within the range of the cluster size are retained throughout the medium~\cite{Rowlands2008,Rowlands2009}. Equation (\ref{Gkspace}) thus reduces to
\begin{equation}
	{\fl} G_{\sigma}(\b{K}_n,\b{k}',i\omega_l)=g_{\sigma}(\b{K}_n,\b{k}',i\omega_l)+g_{\sigma}(\b{K}_n,\b{k}',i\omega_l)\Sigma_{\sigma}(\b{K}_n,i\omega_l)G_{\sigma}(\b{K}_n,\b{k}',i\omega_l),
\end{equation}
where $\b{k}'$ indicates that the momenta $\b{k}$ are restricted to lie within the sublattice BZ associated with each given $\b{K}_n$. Now it is possible to define the ``coarse-grained'' Green's function
\begin{equation}\label{GKn}
	\t{G}_{\sigma}(\b{K}_n,i\omega_l)=\int_{\Omega_{\b{K}_n}}d\b{k}'\,G_{\sigma}(\b{K}_n,\b{k}',i\omega_l),
\end{equation}
which crucially depends only on the set $\{\b{K}_n\}$ since all sublattice momenta $\b{k}'$ have been summed over and hence integrated out of the problem. 

Since the set $\{\b{K}_n\}$ have the same periodicity as a set of ``cluster momenta'' for an isolated real-space cluster of sites with periodic Born-von-Karman boundary conditions imposed (i.e.~a taurus construction), it is possible to view the coarse-grained system as an \emph{effective cluster model}. In real space, this similarly means that the effective medium is now described by the coarse-grained Green's function $\t{G}_{IJ\sigma}(i\omega_l)$ for a set of cluster sites $\{I,J\}$ only. (Throughout this paper, capital letters will be used to denote cluster sites). The cluster momenta and real space cluster sites are related via
\begin{equation}\label{XIJ}
	\frac{1}{N_c}\sum_{n}e^{i(\b{K}.\b{R}_{IJ})}=\delta_{IJ},
\end{equation}
where $N_c$ is the number of sites in the cluster and $n=1,..,N_c$. (Note that for complex unit cells, a separate Fourier transform is required for each sublattice in the unit cell~\cite{Rowlands2014}). Applying (\ref{XIJ}) to $\t{G}_{\sigma}(\b{K}_n,i\omega_l)$ yields
\begin{equation}\label{FIJ}
	{\fl} \t{G}_{IJ\sigma}(i\omega_l)=\frac{1}{N_c}\sum_{n}\int_{\Omega_{\b{K}_n}}d\b{k}'\,\left(i\omega_{l}-\Sigma_{\sigma}(\b{K}_n,i\omega_l)-W(\b{k}')\right)^{-1}e^{i\b{K}_{n}.\b{R}_{IJ}}.
\end{equation}
This implies that the phase factors associated with all sublattice momenta have been neglected, i.e.,~$\exp{(i\b{k}'.\b{R}_{IJ})}\approx{1}$. This is the second main approximation and is consistent with the first.

With a view to defining a self-consistent impurity problem to determine the coarse-grained system, a means of defining an impurity embedded in it is needed. This is achieved in momentum space by first defining a translationally-invariant cavity Green's function ${\cal{G}}$ as follows,
\begin{equation}
	{\fl} \t{G}_{\sigma}(\b{K}_n,i\omega_l)={\cal{G}}_{\sigma}(\b{K}_n,i\omega_l)+{\cal{G}}_{\sigma}(\b{K}_n,i\omega_l)\Sigma_{\sigma}(\b{K}_n,i\omega_l)\t{G}_{\sigma}(\b{K}_n,i\omega_l).
\end{equation}
Fourier-transforming to real space yields
\begin{equation}\label{cavity}
	{\fl} \t{G}_{IJ\sigma}(i\omega_l)={\cal{G}}_{IJ\sigma}(i\omega_l)+\sum_{K,L}{\cal{G}}_{IK\sigma}(i\omega_l)\Sigma_{KL\sigma}(i\omega_l)\t{G}_{LJ\sigma}(i\omega_l).
\end{equation}
An impurity Green's function can now be defined simply by replacing the self-energy with the matrix $\u{V}_{\gamma}$,
\begin{equation}\label{impurity}
	{\fl }G_{IJ\sigma}^{\gamma}(i\omega_l,i\omega_n)={\cal{G}}_{IJ\sigma}(i\omega_l)\delta_{ln}+\sum_{K,m}{\cal{G}}_{IK\sigma}(i\omega_l)V_{\gamma{K}\sigma}(i\omega_l,i\omega_m)G_{KJ\sigma}^{\gamma}(i\omega_m,i\omega_n).
\end{equation}
$\u{V}_{\gamma}$ is diagonal in the cluster site indices and represents a cluster configuration $\gamma$ of impurity one-body potentials, or equivalently, charge and exchange fields $\{\xi_I,\eta_I\}$. In contrast to the self-energy, the one-body potentials are not translationally-invariant in time. Hence it is necessary to know them at all frequency differences relative to the medium frequency $i\omega_l$. By Fourier-transforming (\ref{potential}), the matrix elements of $\u{V}_{\gamma}$ are therefore
\begin{equation}\label{velements}
 {\fl}	v_{\gamma{I}\sigma}(i\omega_l-i\omega_m)=(\epsilon_0-\mu)\delta_{l0}-\frac{1}{2}U\left(i\eta_I(i\omega_l-i\omega_m)+\sigma\xi_I(i\omega_l-i\omega_m)\right)
\end{equation}
for each cluster site $I$ given some cluster impurity field configuration $\gamma$. 

%----------------
\subsubsection{Free energy}\label{secfree}

The effective medium free energy in the dyn-CPA has been derived by considering the free energy for some general configuration of potentials as a fluctuation about the effective medium~\cite{Kakehashi1992}. Then by neglecting the inter-site contributions due to the site-diagonal nature of the self-energy, the fluctuation can be considered in terms of a sum of single-site contributions only. The effective medium free energy is determined by minimising the fluctuations using the CPA~\cite{Kakehashi1992}.

However, this is not a suitable way to derive the required result in the nonlocal CPA since dividing the lattice into non-overlapping real-space cluster contributions breaks the translational invariance of the underlying lattice. In particular, only intra-cluster self-energy and Green's function terms would be considered and inter-cluster terms would not be counted. Instead, a more appropriate way to proceed is to consider only the free energy for fluctuations about the effective cluster model specified by the coarse-grained Green's function (\ref{GKn}) and (\ref{FIJ}). Thus the lattice quantities $g,v,\Sigma,G$ are replaced by the cluster quantities $\u{\cal{G}},\u{V},\u{\Sigma},\u{\t{G}}$.

Hence first consider the free energy for the Hamiltonian $H_{\gamma}$ corresponding to the impurity Green's function (\ref{impurity}),
\begin{equation}\label{freegamma1}
	{\cal{F}}_{\gamma}=-\frac{1}{\beta}\ln\left[\int\prod_{I=1,.,N_c}\delta\xi_I\delta\eta_I\right]e^{-\beta{E}_{\gamma}\left[\left\{\xi_I,\eta_I\right\}\right]}
\end{equation}
with impurity cluster energy functional
\begin{eqnarray}\label{clusterenergy}
     E_{\gamma}\left[\left\{\xi_I,\eta_I\right\}\right]=-\frac{1}{\beta}\ln{Z^1}+\frac{U}{4}\sum_{I,l}\left(|\xi_{I}(i\omega_l)|^2+|\eta_{I}(i\omega_l)|^2\right).
\end{eqnarray}
Replacing the free-electron Green's function in (\ref{lnZ1}) by the cavity Green's function and restricting the trace over the site index to the cluster sites only, the one-body partition function for $H_{\gamma}$ can be manipulated into the form
\begin{equation}\label{lnZ1gamma}
	\ln{Z^1}=\ln{Z^{0'}}+\Tr_I\Tr_{\omega\sigma}\ln[\u{1}-\u{V}_{\gamma}\u{\cal{G}}],
\end{equation}
where for clarity the trace has been split into a trace over the cluster sites denoted by $\Tr_I$ and a trace over spin and frequency denoted by $\Tr_{\sigma\omega}$. Here $Z^{0'}$ is the partition function for the Hamiltonian corresponding to the cavity Green's function. A more useful expression may be obtained by introducing a matrix describing fluctuations from the cluster self-energy,
\begin{equation}
	\u{\delta{v}}=\u{V}_{\gamma}-\u{\Sigma}.
\end{equation}
Now substituting the relation
\begin{equation}
	\ln[\u{1}-\u{V}_{\gamma}\u{\cal{G}}]=\ln[\u{1}-\u{\Sigma}\,\u{\cal{G}}]+\ln[\u{1}-\u{\delta{v}}\,\u{\t{G}}]
\end{equation}
into (\ref{clusterenergy}) yields the expression
\begin{equation}\label{energy}
      E_{\gamma}\left[\left\{\xi_I,\eta_I\right\}\right]={\cal{F}}[\Sigma]+E\left[\left\{\xi_I,\eta_I\right\}\right],
\end{equation}
where the effective cluster free energy is defined by
\begin{equation}\label{Fsigma}
     {\cal{F}}[\Sigma]=-\frac{1}{\beta}\left(\ln{Z^{0'}}+\Tr_I\Tr_{\sigma\omega}\ln[\u{1}-\u{\Sigma}\,\u{\cal{G}}]\right),
\end{equation}
and the energy functional for fluctuations from the self-energy is defined by
\begin{equation}\label{efuncfluc}
  {\fl} E\left[\left\{\xi_I,\eta_I\right\}\right]=-\frac{1}{\beta}\Tr_I\Tr_{\omega\sigma}\ln[\u{1}-\u{\delta{v}}\,\u{\t{G}}]
         +\frac{U}{4}\sum_{I,l}\left(|\xi_{I}(i\omega_l)|^2+|\eta_{I}(i\omega_l)|^2\right).
\end{equation}
Now substituting (\ref{energy}) into (\ref{freegamma1}) yields the final expression,
\begin{equation}\label{freegamma2}
	{\cal{F}}_{\gamma}={\cal{F}}[\Sigma]-\frac{1}{\beta}\ln\int\prod_{I=1,.,N_c}\delta\xi_I\delta\eta_I\,{e}^{-\beta{E}\left[\left\{\xi_I,\eta_I\right\}\right]}.
\end{equation}
Thus the free energy for some impurity field configuration has been rewritten in terms of the free energy for an effective cluster and the free energy for the fluctuation from that effective cluster. The free energy per site is obtained by dividing by $N_c$.

%----------------
\subsubsection{Variational properties}\label{secvariational}

In order to obtain the condition which determines the best possible effective cluster and hence effective medium, the cluster self-energy must minimise the free energy (\ref{freegamma2}). This requires that $\partial{{\cal{F}}_{\gamma}}/\partial{\u{\Sigma}}_{\sigma}(i\omega_l)=0$, where the underscore denotes a matrix in the space of the cluster sites. For convenience, denote $\u{\Sigma}_{\sigma\omega}={\u{\Sigma}}_{\sigma}(i\omega_l)$, $E=E\left[\left\{\xi_I,\eta_I\right\}\right]$ and let
\begin{equation}
	f(E)=\int\prod_{I=1,.,N_c}\delta\xi_I\delta\eta_I\,{e}^{-\beta{E}\left[\left\{\xi_I,\eta_I\right\}\right]}.
\end{equation}
Then the chain rule yields
\begin{equation}\label{chain}
	\frac{\partial{{\cal{F}}_{\gamma}}}{\partial{\u{\Sigma}}_{\sigma\omega}}  =  \frac{\partial{{\cal{F}}}[\Sigma]}{\partial{\u{\Sigma}_{\sigma\omega}}} 
                                                                                                            -\frac{1}{\beta}\frac{\partial}{\partial{E}}[\ln(f(E))]\frac{\partial{E}}{\partial{\u{\Sigma}}_{\sigma\omega}} \nonumber\\
																																						                                           =  \left< \frac{\partial{{\cal{F}}}[\Sigma]}{\partial{\u{\Sigma}}_{\sigma\omega}}+\frac{\partial{E}}{\partial{\u{\Sigma}}_{\sigma\omega}} \right>,
\end{equation}
where
\begin{equation}\label{weight}
	 \left<\sim\right>=\frac{\displaystyle\int\!\!\prod_{I=1,.,N_c}\!\!\!\delta\xi_{I}\delta\eta_{I}\,\left(\sim\right)e^{-\beta{E}\left[\left\{\xi_{I},\eta_{I}\right\}\right]}}{\displaystyle\int\!\!\prod_{I=1,.,N_c}\!\!\!\delta\xi_{I}\delta\eta_{I}\,\,e^{-\beta{E}\left[\left\{\xi_{I},\eta_{I}\right\}\right]}}
\end{equation}
is a thermal average which arises from taking the functional derivative of $\ln(f(E))$. Note that ${\cal{F}}[\Sigma]$ is independent of the thermal average. Inserting (\ref{Fsigma}) and (\ref{efuncfluc}) into (\ref{chain}), taking the functional derivatives and utilising cyclic permutations of the matrices under the trace operation yields
\begin{equation}\label{variation}
	\frac{\partial{{\cal{F}}_{\gamma}}}{\partial{\u{\Sigma}}_{\sigma\omega}}=-\frac{1}{\beta} \left\{ \Tr_{I}\left[\u{\t{G}}\left<\u{t}_{\gamma}\right>\u{\t{G}}\right]_{\sigma\omega}
                                                                                                             -\Tr_{I}\Tr_{\sigma\omega}\left<\u{t}_{\gamma}\right> \frac{\partial{{\cal{F}}}[\Sigma]}{\partial{\u{\Sigma}_{\sigma\omega}}} \right\}.
\end{equation}
Here $\u{t}_{\gamma}$ is the cluster scattering t-matrix for the impurity cluster configuration $\gamma$ defined by
\begin{equation}
	\u{t}_{\gamma}=\left[\u{1}-\u{\delta{v}}\,\u{\t{G}}\right]^{-1}\u{\delta{v}}.	
\end{equation}
According to (\ref{variation}), in order that $\partial{{\cal{F}}_{\gamma}}/\partial{\u{\Sigma}}_{\sigma\omega}=0$, the condition
\begin{equation}
	\left<\u{t}_{\gamma}\right>=0
\end{equation}
must be satisfied. This can be straightforwardly manipulated into the form
\begin{equation}\label{selfconsistency}
	\left<{G}_{IJ\sigma}^{\gamma}(i\omega_{l},i\omega_{l})\right>=\t{G}_{IJ\sigma}(i\omega_{l}),
\end{equation}
where the thermal weights to be used when performing the cluster averaging are given by (\ref{weight}). The diagonal frequency matrix elements of the impurity Green's function are defined by
\begin{equation}\label{impurity2}
	G_{IJ\sigma}^{\gamma}(i\omega_l,i\omega_l)=\left[\left(\u{\t{G}}^{-1}-\u{\delta{v}}\right)^{-1}\right]_{Il{\sigma}Jl{\sigma}},
\end{equation}
which is obtained by combining (\ref{cavity}) and (\ref{impurity}). Here the underscore denotes a matrix in the cluster-site, frequency and spin indices. Note that since $\u{\delta{v}}$ in the RHS of (\ref{impurity2}) is not-diagonal in the frequency indices, this involves the inversion of a large matrix before the diagonal frequency elements are taken.  

In practice the result (\ref{selfconsistency}) can be achieved computationally by using a self-consistent algorithm. Self-consistency is achieved when the LHS of (\ref{selfconsistency}) is given by (\ref{impurity2}) with thermal weights (\ref{weight}) and the RHS by (\ref{FIJ}). The functional integrals in (\ref{weight}) are defined by
\begin{equation}\label{funcint2}
	 \int\delta\xi_I=\int\sqrt{\frac{\beta{U}}{4\pi}}d\xi_I(0)\left[\prod_{l=1}^{\infty}\frac{\beta{U}}{2\pi}\,d\,\mathrm{Re}\,\xi_I(i\omega_l)\,d\,\mathrm{Im}\,\xi_I(i\omega_l)\right],
\end{equation}
where $d\xi_I(0)$ is the zero-freqency component, and these may be evaluated using a quantum cluster impurity solver (see section~\ref{secevaluation}).

%----------------
\subsubsection{Medium Green's function}\label{secmedium}

The cluster self-energy specifies the lattice or medium self-energy as a step function in k-space centred at the cluster values. Furthermore, the medium Green's function and density of states (DOS) at the cluster sites are equal to those of the effective cluster model. Therefore effective medium properties can be calculated from knowledge of the cluster charge-field and exchange-field matrices for some given field configuration $\gamma$. These are defined by
\begin{equation}\label{ssfcharge}
	\u{\eta}^{\gamma}=\frac{1}{\beta}\sum_{\omega\sigma}\u{G}^{\gamma}_{\omega\sigma} \;\; \mathrm{and} \;\;\; \u{\xi}^{\gamma}=\frac{1}{\beta}\sum_{\omega\sigma}\sigma\u{G}^{\gamma}_{\omega\sigma}
\end{equation}
respectively, and are matrices in the cluster-site index only, while $\u{G}^{\gamma}$ is a matrix in the cluster-site, spin and frequency indices. The average charge and magnetization per site are defined by the thermal averages
\begin{equation}\label{ssfavgcharge}
	n=\left<\u{\eta}^{\gamma}_{I}\right> \;\; \mathrm{and} \;\;\; m=\left<\u{\xi}^{\gamma}_{I}\right>
\end{equation}
respectively, where any cluster-site matrix element $I$ can be chosen since translational invariance means they all have the same value on the average (but not independently from the other sites as in the dyn-CPA).

On the other hand, it is known that the step function form of the self-energy leads to non-unique results for small cluster sizes~\cite{Rowlands2006b,Rowlands2008}. In reality the self-energy should be a smooth function for any cluster size $N_c$~\cite{Rowlands2008,Rowlands2009} but the step function form is used as it has been proven to preserve the analytic properties of the Green's function for a finite cluster size~\cite{Hettler2000}. In DCA calculations the self-energy is often smoothed out after the algorithm has converged by using a spline through the $\{\b{K}_n\}$ values~\cite{Maier2002}. However, this means that the medium Green's function is no longer equal to that of the effective cluster model. Moreover, in references~\cite{Rowlands2008,Rowlands2009} it was argued that the self-energy should not be interpolated because this can introduce new physics not obtained through the calculation and is not guaranteed to preserve the analytic properties of the Green's function. As a result, a reformulation of the nonlocal CPA was developed~\cite{Rowlands2008} based on the observation that (\ref{XIJ}) is in fact a special case of the more general relation
\begin{equation}\label{phase}
	\frac{1}{N_c}\sum_{n}e^{i(\b{K}.\b{R}_{IJ}-\phi)}=\delta_{IJ},
\end{equation}
which indicates that the set $\{\b{K}_n\}$ must shift in k-space to cancel the arbitrary phase $\phi$. Hence there are many possible sets of cluster momenta which satisfy (\ref{XIJ}). In addition to averaging the impurity Green's function over disorder configurations, the reformulation~\cite{Rowlands2008} also self-consistently averages the impurity Green's function over possible choices of $\{\b{K}_n\}$. The theory is guaranteed to be analytic and was shown to systematically reduce the magnitude of the discontinuities in the self-energy steps, hence converging towards a unique result. Importantly, since the medium Green's function is determined self-consistently, physical properties are calculated directly from the medium Green's function. For simplicity, in the present paper the dyn-NLCPA has been derived using only one set of cluster momenta although extension to multiple sets is straightforward.

%----------------
\subsubsection{Evaluation of the functional integrals}\label{secevaluation}

Kakehashi~\cite{Kakehashi1992} applied the QMC method to calculate the functional integrals (\ref{funcint2}) for a single-site impurity in the dyn-CPA. An alternative method is the harmonic approximation~\cite{Kakehashi2002a,Tamashiro2011} which is an analytic approach based on neglecting mode-mode couplings and is exact up to order $U^2$ with the advantage of being computationally less demanding. However, the QMC method as applied by Kakehashi can be straightforwardly extended to the case of a cluster of sites in the dyn-NLCPA. 

First the interval $[0,\beta]$ in the imaginary time representation is discretized into $N$ mesh points $\tau_n$, each of duration $\Delta\tau=\beta/N$ such that $\tau_n=n\Delta\tau$. This means all matrix elements involving time need to be multiplied by $\Delta\tau$ and hence the matrices have dimension $N{\times}N$ in the time index. The functional integrals in the imaginary time representation are defined by (\ref{funcinttime}) and are replaced by repeated $N^{'}_{\xi_I}$-fold integrals for each cluster site. A simple approximation is to only consider the exchange fields (the one-field method) and to deal with the charge fields by demanding charge neutrality for the cluster. Then the thermal average of the impurity Green's function (\ref{selfconsistency}) is calculated as
\begin{equation}
{\fl} \left(\u{G}^{\gamma}\right)_{\tau_1\tau_2}=\frac{\displaystyle \int\left[\prod_{I=1}^{N_c}\prod_{n=1}^{N^{'}_{\xi_I}}d\xi_I(\tau'_n)\right]  
                                                                                       			 \left(\u{G}^{\gamma}\right)_{\tau_1\tau_2} \det\left[\u{1}-\u{\delta{v}}\,\u{\t{G}}\right] 
                                                                                        					\exp{\left[-\frac{U\Delta\tau'}{4\beta}\sum_{i=1}^{N_c}\sum_{n=1}^{N^{'}_{\xi_I}}\xi_I(\tau'_n)^{2}\right]}  }
																																																							{\displaystyle\int\left[\prod_{I=1}^{N_c}\prod_{n=1}^{N^{'}_{\xi_I}}d\xi_I(\tau'_n)\right] 
                                                                                         \det\left[\u{1}-\u{\delta{v}}\,\u{\t{G}}\right]\exp{\left[-\frac{U\Delta\tau'}{4\beta}\sum_{i=1}^{N_c}\sum_{n=1}^{N^{'}_{\xi_I}}\xi_I(\tau'_n)^{2}\right]}},
\end{equation}
where the underscore denotes a matrix in the cluster-site and spin indices. Here $\Delta\tau'=\beta/N^{'}_{\xi_I}$ and $\tau'_{n}=n\Delta\tau'$, and the fraction $\Delta\tau'$ is taken independently of the time interval $\Delta\tau=\beta/N$ which defines the size of the time aspect of the matrices. The sampling can be taken to be Gaussian.

After obtaining suitable initial inputs for the chemical potential and self-energy by solving the charge-neutrality condition for the cluster via (\ref{ssfavgcharge}), the impurity Green's function $\left({G}_{IJ\sigma}^{\gamma}\right)_{\tau_n{0}}$ is calculated at each Monte Carlo time step by solving the $N{\times}N$ simultaneous linear equations with Gaussian random variables generated by the polar method. When the dyn-NLCPA self-consistency condition (\ref{selfconsistency}) expressed as $\left<\left({G}_{IJ\sigma}^{\gamma}\right)_{\tau_n{0}}\right>=(\t{G}_{IJ\sigma})_{\tau_n{0}}$, $(n=0,1,\cdots,N-1)$ is not satisfied after the Monte Carlo evaluation, a new guess for the self-energy is obtained by solving (\ref{cavity})
and the dyn-NLCPA equations can then be solved self-consistently. Self-consistency is achieved when both the dyn-NLCPA equations and charge-neutrality are satisfied.

During the algorithm, the relevant time-dependent quantities need to be diagonalised in the frequency representation. For example, the self-energy appearing in (\ref{FIJ}) can be calculated via the Fourier transform
\begin{equation}
	\Sigma_{\sigma}(\b{K}_n,i\omega_l)=\sum_{n=1}^{N/2}\Sigma_{\sigma}(\b{K}_n,\tau_n)e^{i\omega_l\tau_n}\Delta\tau,
\end{equation}
where the Matsubara frequencies are $\omega_l=(2l+1)\pi/\beta$, with $l$ an integer such that $-N/2+1{\le}2l+1{\le}N/2$. Furthermore, using the relation for the free-electron integrated DOS,
\begin{equation}
	\int{d\varepsilon}\;\rho_{0}(\varepsilon)=\int{d\varepsilon}\;\frac{1}{N}\sum_{\b{k}}\delta(\varepsilon-W(\b{k}))
\end{equation}
for real energies $\varepsilon$, equation (\ref{FIJ}) for the coarse-grained Green's function can be rexpressed in the form
\begin{equation}\label{lehmann}
	\t{G}_{IJ\sigma}(i\omega_l)=\sum_{n}\int{d}\varepsilon\,\frac{\rho_0(n,\varepsilon)}{i\omega_l-\Sigma_{\sigma}(\b{K}_n,i\omega_l)-\varepsilon}\exp{(i\b{K}_{n}.\b{R}_{IJ})},
\end{equation}
where a free-electron sublattice integrated DOS is defined by
\begin{equation}
	\int{d\varepsilon}\;\rho_{0}(n,\varepsilon)=\int{d\varepsilon}\;\frac{N_c}{N}\sum_{\b{k}'\in\Omega_{\b{K}_n}}\delta(\varepsilon-W(\b{k}'))
\end{equation}
for each sublattice $n$. Note that the set $\{\rho_{0}(n,\varepsilon)\}$ may be specified by a model DOS or, for realistic applications, calculated via first-principles methods.

%-------------------------------------------------------------------
\subsection{Static approximation}\label{static}

In the static approximation the imaginary-time dependence of the charge and exchange fields is neglected and only their zero-frequency components are retained, $\xi_I(0)=\beta^{-1}\int_0^\beta\xi_I(\tau)d\tau$, and $\eta_I(0)=\beta^{-1}\int_0^\beta\eta_I(\tau)d\tau$ (see , for example, reference~\cite{Fehske1984} for a discussion of the validity of the static approximation in the spin-fluctuation theory). It follows that the one-body potentials (\ref{velements}) now also only have zero frequency components given by
\begin{equation}\label{vstatic}
 v_{\gamma{I}\sigma}(0)=(\epsilon_0-\mu)-\frac{1}{2}U\left(i\eta_I(0)+\sigma\xi_I(0)\right)
\end{equation}
for each cluster site $I$ and field configuration $\gamma$. The impurity problem represented by (\ref{selfconsistency}) reduces to a static thermal average over the possible field configurations and the functional integrals (\ref{funcint2}) reduce to ordinary integrals,
\begin{equation}
	\int\delta\xi_I\rightarrow\int\!\sqrt{\frac{\beta{U}}{4\pi}}\,d\xi_I \;\;\; \textrm{and} \;\; \int\delta\eta_I\rightarrow\int\!\sqrt{\frac{\beta{U}}{4\pi}}\,d\eta_I.
\end{equation}
Hence the thermal weights (\ref{weight}) reduce to
\begin{equation}\label{weight2}
	 \left<\sim\right>=\frac{\displaystyle\int\!\!\prod_{I=1,.,N_c}\!\!\!d\xi_{I}d\eta_{I}\,\left(\sim\right)e^{-\beta{E}\left[\left\{\xi_{I},\eta_{I}\right\}\right]}}{\displaystyle\int\!\!\prod_{I=1,.,N_c}\!\!\!d\xi_{I}d\eta_{I}\,\,e^{-\beta{E}\left[\left\{\xi_{I},\eta_{I}\right\}\right]}},
\end{equation}
and the energy functional (\ref{efuncfluc}) for fluctuations from the self-energy becomes
\begin{equation}\label{Estatic}
	E(\{\xi_I,\eta_I\})=-\frac{1}{\beta}\Tr_I\Tr_{\omega\sigma}\ln[\u{1}-\u{\delta{v}}(0)\u{\t{G}}]+\frac{U}{4}\sum_{I}(\xi_I^2+\eta_I^2).
\end{equation}
Here the underscore denotes a matrix in the cluster-site, spin and frequency indices. For calculational purposes, the relation $\mathrm{Tr(ln)=ln(det)}$ can be used to trade the trace over cluster site and/or spin for a determinant.

For $N_c>1$ the above is a nonlocal generalization of the SSF theory. The values for the field configurations should ideally be obtained via Monte Carlo sampling. A simpler alternative is to sample only the cluster exchange-field configurations $\{\xi_I\}$ and to associate with them the corresponding set of saddle point charges $\{\eta_I^{*}\}$ such that overall charge neutrality is maintained for the cluster (and hence per site due to translational invariance). Then $\gamma$ represents a cluster exchange-field configuration only and the integrals over the charge fields disappear from the thermal weights so that (\ref{weight2}) further reduces to
\begin{equation}\label{weight3}
	 \left<\sim\right>=\frac{\displaystyle\int\!\!\prod_{I=1,.,N_c}\!\!\!d\xi_{I}\,\left(\sim\right)e^{-\beta{E}\left[\left\{\xi_{I},\eta_I^{*}\right\}\right]}}{\displaystyle\int\!\!\prod_{I=1,.,N_c}\!\!\!d\xi_{I}\,\,e^{-\beta{E}\left[\left\{\xi_{I},\eta_I^{*}\right\}\right]}}.
\end{equation}
The set $\{\eta_I^{*}\}$ are those values which minimise the functional $E[\{\xi_I,\eta_I\}]$ for the given cluster exchange-field configuration $\gamma=\{\xi_I\}$, and can be obtained by ensuring the charge fields appearing in $V_{\gamma}$ are consistent with the charge fields obtained from (\ref{ssfcharge}). Note that for $N_c>1$, charge is allowed to transfer between the cluster sites when calculating $\{\eta_I^{*}\}$. An example algorithm for the saddle-point approximation is:
\begin{enumerate}
	\item{Make a guess for $\mu$ for the desired electron filling.} 
	\item{Make a guess for the cluster self-energy $\u{\Sigma}_{\sigma}(i\omega_l)$ for all frequencies.}
	\item{Calculate the coarse-grained Green's function $\u{\t{G}}_{\sigma}(i\omega_l)$ for all frequencies via (\ref{FIJ}).}
	\item{Calculate the cavity Green's function $\u{\cal{G}}_{\sigma}(i\omega_l)$ by solving the matrix equation (\ref{cavity}).}
	\item{For each configuration $\gamma$ of exchange-field values at the cluster sites $\{\xi_I\}$:}
		\begin{enumerate}[(a)]
			\item{Make a guess for the corresponding set of saddle-point charges $\{\eta_I^{*}\}$. (Note that $\{\eta_I^{*}\}$ are themselves complex at the saddle-points).}
			\item{Calculate the cluster potential matrix $\u{V}_{\gamma}$ via (\ref{vstatic}) and hence the impurity Green's function $\u{G}_{\sigma}^{\gamma}(i\omega_l)$ via (\ref{impurity2}) with $\u{\delta{v}}=\u{\delta{v}}(0)$.}
			\item{Recalculate the set of saddle-point charges $\{\eta_I^{*}\}$ via (\ref{ssfcharge}) and repeat from (b) until self-consistency is achieved for $\{\eta_I^{*}\}$.}
		\end{enumerate}
	\item{Calculate the energy functional $E[\{\xi_I,\eta_I^{*}\}]$ via (\ref{Estatic}) for each configuration $\gamma$. (Note this requires knowledge of $\u{\t{G}}_{\sigma}(i\omega_l)$ over all frequencies).}
	\item{Calculate the thermal average (\ref{selfconsistency}) using weights (\ref{weight3}) to give a new guess for $\u{\t{G}}_{\sigma}(i\omega_l)$.}
	\item{Calculate a new guess for $\u{\Sigma}_{\sigma}(i\omega_l)$ for all frequencies by solving (\ref{cavity}) using $\u{\cal{G}}_{\sigma}(i\omega_l)$ from step (iv). Repeat as necessary from step (ii) until self-consistency for $\u{\Sigma}_{\sigma}(i\omega_l)$ is achieved.}
	\item{Check if the average charge per site (\ref{ssfavgcharge}) is consistent with the desired electron filling. Repeat as necessary from step (i) using a new guess for $\mu$ until the desired filling is obtained.}
\end{enumerate}

%--------------------------------------------------------------------------------------------------------------------------------------------

\section{Results}\label{results}

\begin{figure}
 \begin{center}
 \scalebox{1.0}{\i{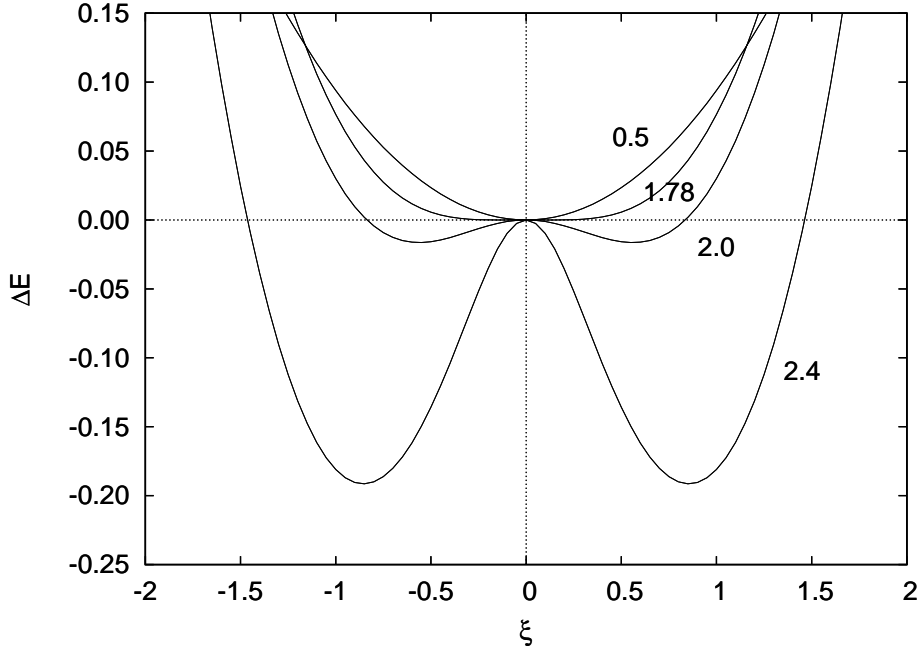}}
 \caption{\label{ssf} Energy potential curves $\Delta{E}(\xi)=E(\xi)-E(0)$ for the 1D Hubbard model calculated at half-filling using the single-site SSF. Results are shown for $U/W$ = 0.5, 1.78, 2.0, and 2.4 in the paramagnetic regime at $T/W$ =0.06, where $W$ denotes the half-bandwidth. The curve just dips below the zero axis at $U=1.78$, indicating the formation of the local moment state. }
 \end{center}
\end{figure}

\begin{figure}
 \begin{center}
\addtolength{\columnsep}{-10mm}
{\renewcommand{\arraystretch}{-100}
 \begin{tabular}{c} 
 \scalebox{0.8}{\i{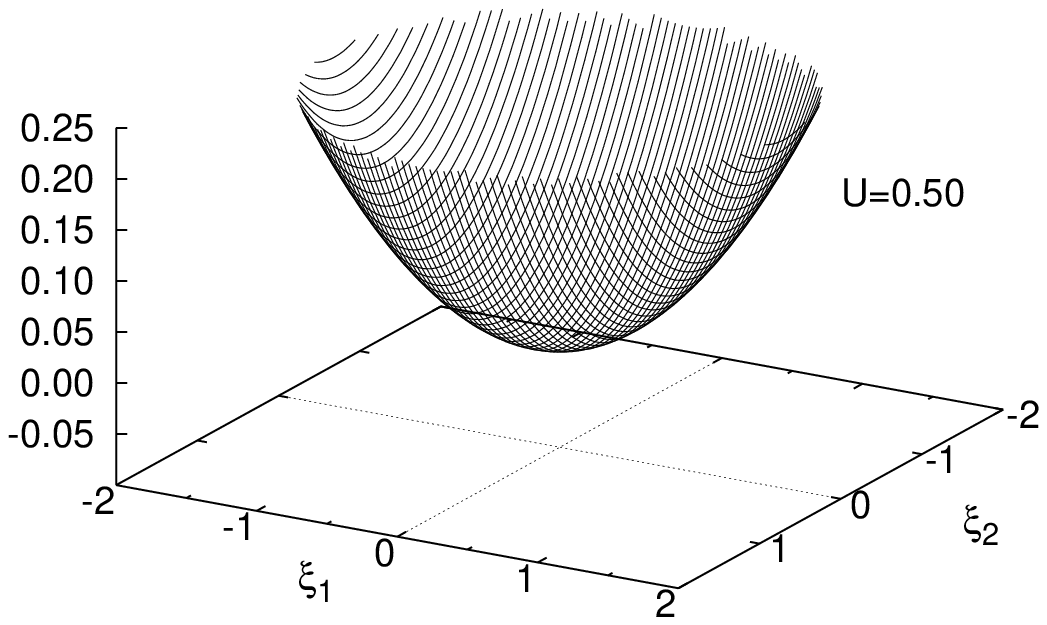}} \\
 \scalebox{0.8}{\i{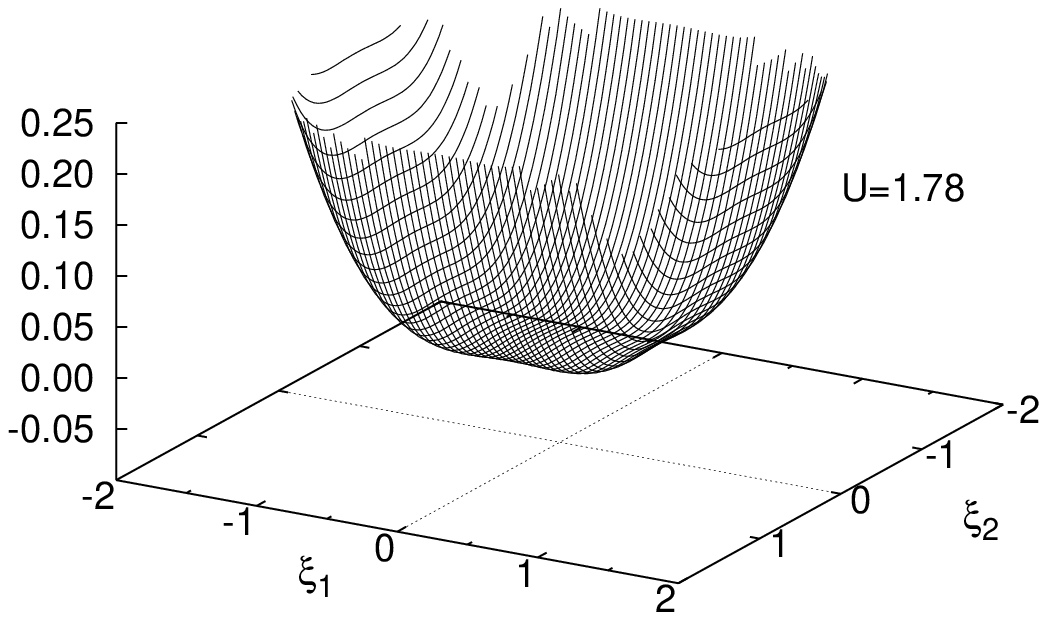}} \\
 \scalebox{0.8}{\i{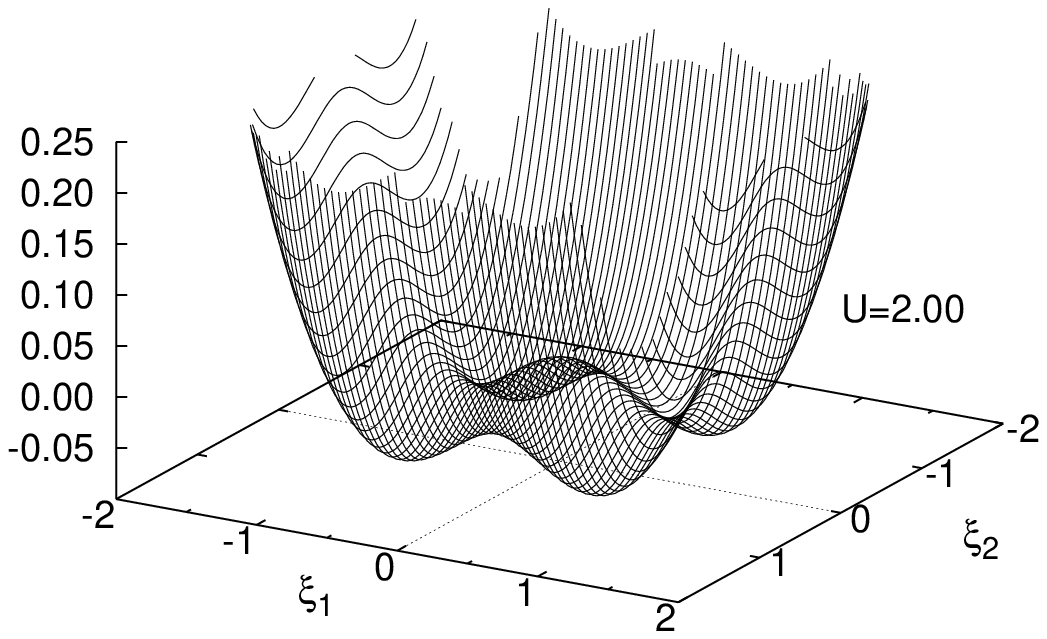}} 
 \end{tabular}}
 \caption{\label{3d} Energy potential surfaces $\Delta{E}(\xi_1,\xi_2)=E(\xi_1,\xi_2)-E(0,0)$ for the 1D Hubbard model calculated at half-filling with $N_c=2$. Results are shown for $U/W$ = 0.5, 1.78, 2.0 in the paramagnetic regime at $T/W$ = 0.06, where $W$ denotes the half-bandwidth. The surface plot has already dipped below the zero plane in the $U$ =1.78 calculation, indicating the formation of the local moment state.}
\end{center}
\end{figure}

\begin{figure}
 \begin{center}
\addtolength{\columnsep}{-10mm}
{\renewcommand{\arraystretch}{-100}
 \begin{tabular}{c} 
 \scalebox{0.7}{\i{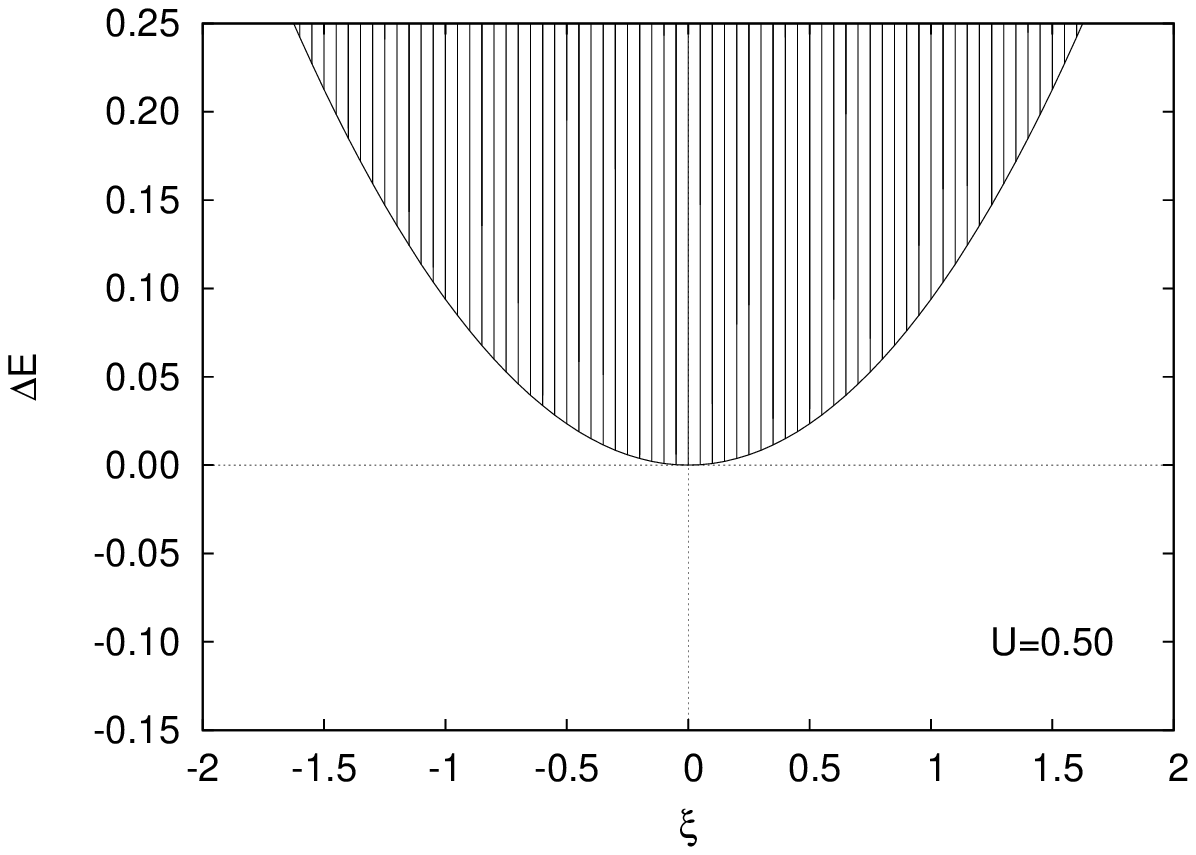}} \\
 \scalebox{0.7}{\i{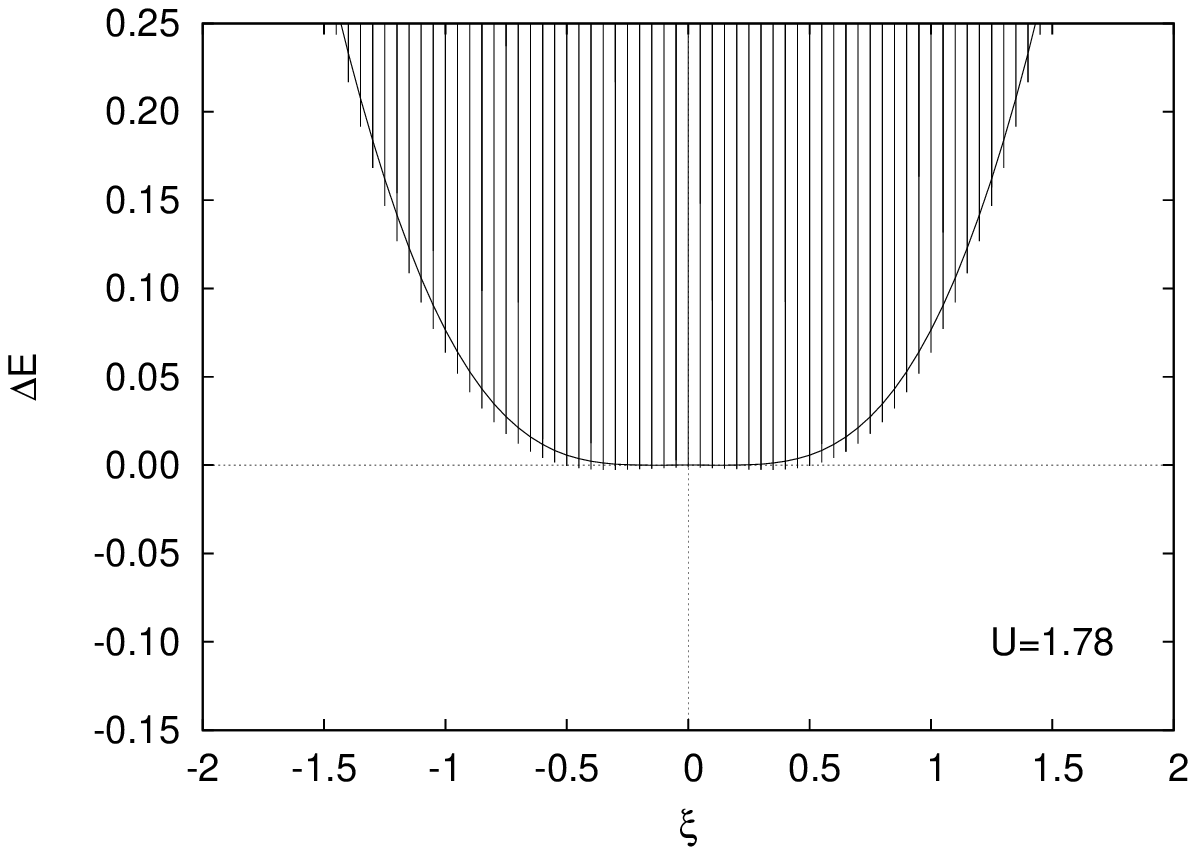}} \\
 \scalebox{0.7}{\i{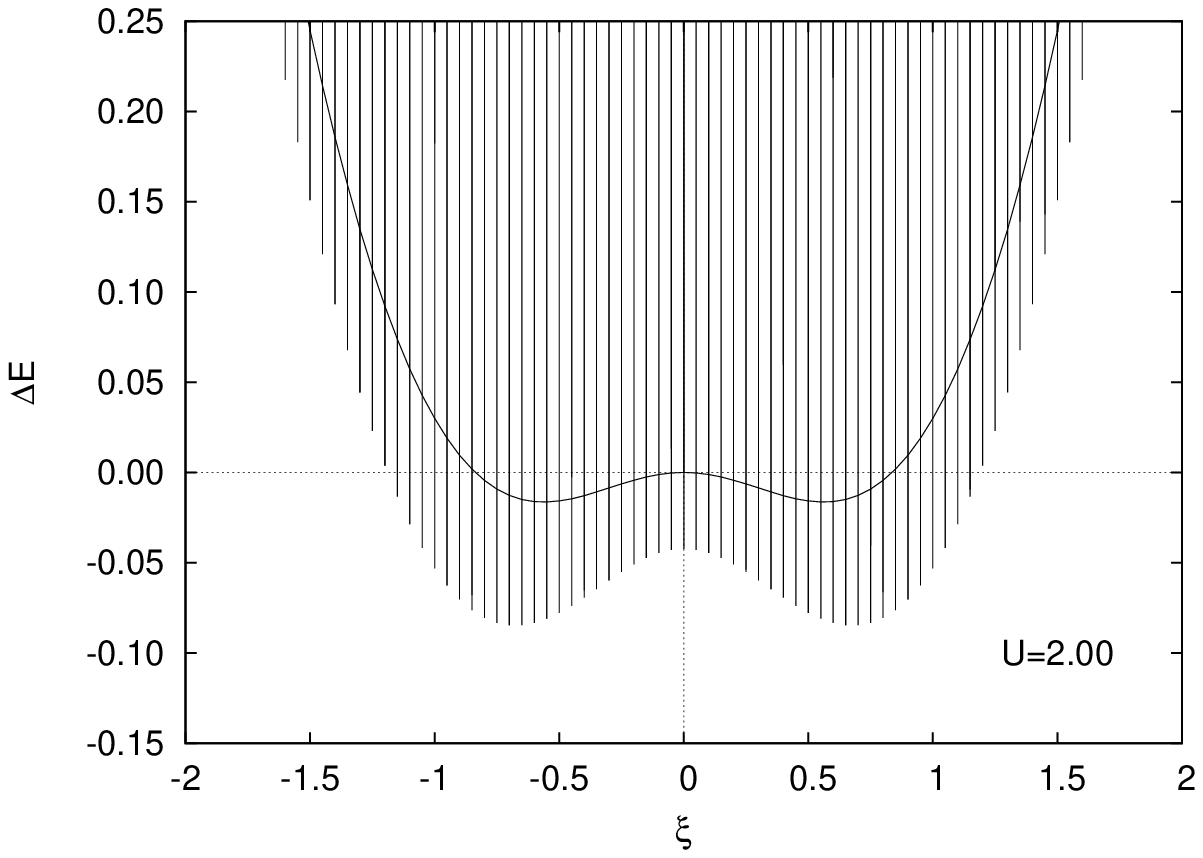}}
 \end{tabular}}
 \caption{\label{nlssf} Energy potential surfaces $\Delta{E}(\xi_1,\xi_2)=E(\xi_1,\xi_2)-E(0,0)$ of figure~\ref{3d} projected onto one axis for the 1D Hubbard model calculated at half-filling with $N_c=2$. Results are shown for $U/W$ = 0.5, 1.78, 2.0 in the paramagnetic regime at $T/W$ = 0.06, where $W$ denotes the half-bandwidth. The surface projections are shown by the shaded areas with vertical lines and the $N_c=1$ results are shown by the solid curves. The surface projection has already dipped below the zero axis in the $U/W$ = 1.78 calculation, indicating the formation of the local moment state.}
\end{center}
\end{figure}

\begin{figure}
 \begin{center}
 \scalebox{0.8}{\i{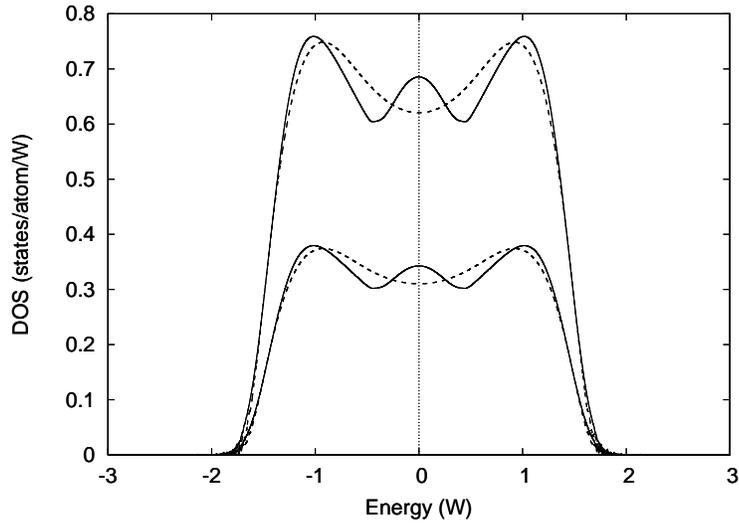}}
 \caption{\label{dos} Density of states (DOS) per site for the model of figure~\ref{nlssf} with $U/W$ = 1.78. The solid lines show the spin-resolved and total DOS for the $N_c=2$ calculation, and the dashed lines show the spin-resolved and total DOS for $N_c=1$.}
 \end{center}
\end{figure}

\begin{figure}
 \begin{center}
 \scalebox{0.8}{\i{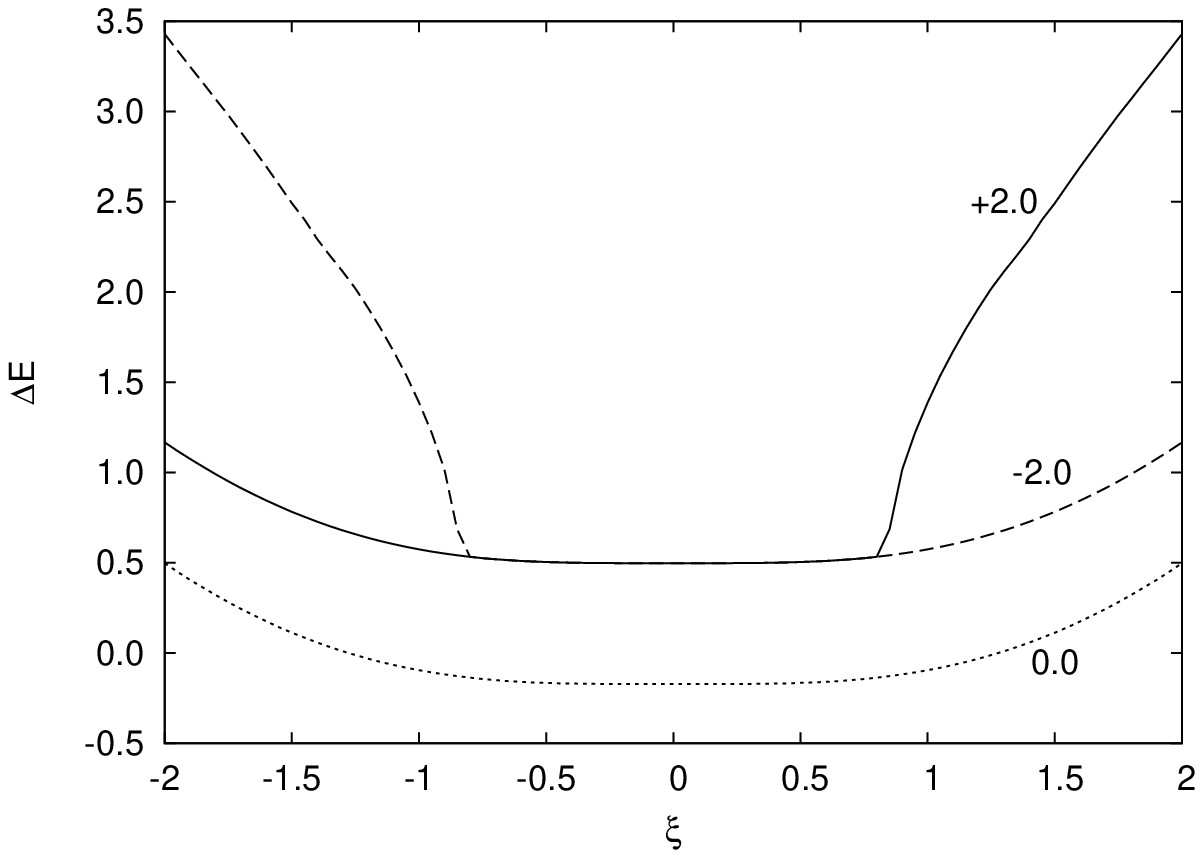}}
 \caption{\label{enpart} Energy potential curves $\Delta{E}(\xi_1,\xi_2=\alpha)=E(\xi_1,\alpha)-\o{E}(0,0)$ for the model of figure~\ref{nlssf} with $U=1.72$ calculated with a cluster size $N_c=2$. Results are shown for the specific configurations where $\xi_1$ is varied but $\xi_2$ is kept fixed at the values $\alpha$ = +2.0, 0.0, and -2.0 as indicated by the solid, dotted, and dashed curves respectively.}
 \end{center}
\end{figure}

As a first step in understanding the implications of the new theory, calculations are presented here within the static approximation to the dyn-NLCPA as detailed in section~\ref{static}. The aim is to study the effect of nonlocal spatial correlations on the formation of the ``local moment'' state in the absence of dynamical correlations. For this purpose, the simplest possible model was chosen, i.e.~the one-dimensional Hubbard model with nearest-neighbour hopping. A cluster size $N_c$ = 2 was used which completely takes into account nearest-neigbour correlations in the field configurations, and the band was half-filled. The frequency sums in (\ref{Estatic}) were transformed into integrals along the real axis with imaginary part $10^{-3}$, and a simple uniform sampling of the exchange field configurations $\{\xi_I\}$ was used with a mesh value $0.05$ on both cluster sites. The conventional set of cluster momenta was used so that $\b{K}_1=-\pi/2a$, $\b{K}_2=+\pi/2a$.

In order to illustrate the physics of local moment formation in the paramagnetic regime, figure~\ref{ssf} shows the energy potential curves calculated using the single-site SSF (i.e.~as defined by (\ref{Estatic}) with $N_c$ = 1) at a relatively low temperature $T/W$ = 0.06, where $W$ is the half-bandwidth which was taken to be the energy unit~\cite{Kakehashi2012}. The origin has been shifted by subtracting the value $E(\xi=0)$. For small $U$ it can be seen that the energy potential curve has a single minimum at $\xi=0$. When $U$ is sufficiently large, the ``local moment'' state appears as a double minimum. At temperature $T/W=0.06$, the critical value for the appearance of the local moment state is calculated to be $U$ = 1.78. This is defined as the point where the curve first dips below the zero axis. In the two-field form of the interaction (\ref{interaction}), a double minimum indicates that there are two preferred directions (and magnitudes) for the exchange fields. The total spin per atom can thermally fluctuate between the two minima. As $U$ becomes large, these spin fluctuations which self-consistently maintain the exchange fields can have a large magnitude as evident for the example value $U=2.4$. However, the energy potential curves are symmetric about the z-axis when the system is in the paramagnetic regime so that the overall magnetization is zero. 

In the intermediate and low temperature regimes, the physics of itinerant electron magnetism is usually dominated by dynamical fluctuations rather than the thermal spin fluctuations and often leads to the disappearance of the double minimum ``local moment'' state~\cite{Kakehashi2012,Kakehashi2002a}. Hence for a given value $U$, dynamical correlations usually delay the onset of the local moment state as the temperature is lowered. Since the formation of the local moment state is a precursor to magnetic ordering, this in turn leads to a lowering of the Curie temperature as expected from experiment. Here in the absence of dynamical correlations, the effect of nonlocal spatial correlations on the formation of the local moment state is now investigated. 

Figure~\ref{3d} shows results for the same model described above using a cluster size $N_c$ = 2. Since the field configurations $\{\xi_I\}$ can vary independently on both sites, the energy potential can be plotted as a surface in 3D with $\xi_1$ plotted along the x-axis and $\xi_2$ along the y-axis. In order to aid comparison with the $N_c$ = 1 calculations, a projection of the 3D plots were taken along one axis and plotted together with the $N_c$ = 1 results as shown in figure~\ref{nlssf}. For small $U$, the projected energy potential curve is very similar to the $N_c$ = 1 result. As $U$ increases, some differences appear and the local moment state occurs at a smaller critical value of $U$ = 1.72. Indeed, it can be seen in figure~\ref{nlssf} that when the local moment state just appears in the $N_c$ = 1 calculation at $U$ = 1.78 shown by the solid curve, the $N_c$ = 2 curve already shows better-defined local minima. Furthermore, figure~\ref{dos} shows that a central peak has emerged in the DOS at this value of $U$. (Note that a similar peak was recently observed in Hubbard III calculations using the nonlocal-CPA~\cite{Rowlands2014}). At $U$ = 2.00, the $N_c$ = 2 curve in figure~\ref{nlssf} shows a much deeper minimum with well-defined local moments. These features arise from the fact that in the $N_c$ = 2 calculation, each impurity cluster site feels the effect of individual field configurations from the other cluster site as well as the surrounding effective medium. To investigate this further, figure~\ref{enpart} shows energy potential curves for the $N_c$ = 2 critical value $U$ = 1.72 plotted with the exchange-field value on site one $\xi_1$ varied while that on site two is fixed at $\xi_2$ = +2.0, 0, or -2.0 respectively. It is evident that the deeper ``well''-type features which lead to a double minimum arise from configurations where the nearest-neighbour exchange field has a large magnitude in the $+z$ and $-z$ directions. On the other hand, the smoother features of the curve which appear similar to the $N_c$ = 1 calculation arise from configurations where the nearest-neighbour exchange field value is small.

Next, the Coulomb interaction was kept fixed at the $N_c$ = 1 critical value $U$ = 1.78 and the temperature was varied. At high temperatures, there is no observable difference between the $N_c$ = 1 and $N_c$ = 2 calculations. For example, if the Couloumb interaction $U$ = 2.00 is chosen and the temperature varied, the local moment state appears at $T/W$ = 0.33 for both $N_c$ = 1 and $N_c$ = 2. However, as the temperature is lowered, it is found that the local moment state appears at the higher temperature, $T/W$ = 0.226 in the $N_c$ = 2 calculation compared to 0.06 for the $N_c$ = 1 calculation. This finding compliments the above one at fixed temperature, both indicating that nearest-neighbour nonlocal correlations in the static approximation stabilise the local moment state.

Hence like recent static Hubbard III calculations investigating the metal-insulator transition using the nonlocal CPA~\cite{Rowlands2014}, the behaviour is opposite to what would be expected. In the case of the Hubbard III calculations, the reason for such behaviour appears to be that the nonlocal-CPA cannot describe electron localization and hence the ``improved'' static configurational averaging always favours the metallic state~\cite{Ekuma2014}. In the present context of itinerant electron magnetism, the same type of average employed by the dyn-NLCPA in the static approximation encourages the formation of the local moment state at a higher temperature, in contrast to dynamical theories which include a description of nonlocal spatial correlations such as the DCA. This result indicates that involving dynamical correlations is essential if a description of nonlocal spatial correlations is included using the nonlocal CPA.

%--------------------------------------------------------------------------------------------------------------------------------------------
\section{Conclusions}\label{conclusions}

A dynamical nonlocal coherent potential approximation (dyn-NLCPA) has been derived within the functional integral approach to the interacting electron system. The theory improves the description of itinerant electron magnetism by including nonlocal correlations in the exchange field configurations. The free energy is proven to be variational with respect to the effective medium and hence the theory is thermodynamically-consistent. The variational requirement yields a self-consistency condition involving a thermal average of the Green's function for a cluster of impurity field configurations. 

As a first step in understanding the implications of the theory, results have been presented for a simple model within the static approximation to the dyn-NLCPA in order to investigate the effects of nonlocal spatial correlations in the absence of dynamical correlations. The calculations reveal that in this case the inclusion of nearest-neigbour correlations between the field configurations modifies the shape of the energy potential curve in the paramagnetic regime and aids the formation of the local moment state. Hence consistent with recent static Hubbard III calculations using the nonlocal CPA~\cite{Rowlands2014}, inclusion of nonlocal spatial correlations alone gives an opposite effect to the case where both nonlocal spatial correlations and dynamical correlations are involved~\cite{Ekuma2014}. This result indicates that involving dynamical correlations is essential if a description of nonlocal spatial correlations is included using the nonlocal CPA.

Unlike the implementation within the static approximation, the dyn-NLCPA implemented with a QMC cluster impurity solver is expected to lower the Curie temperatures for simple metals compared to the single-site dyn-CPA and obtain magnetization curves closer to the experimental values. Implementation of the full dynamical theory with calculations for realistic systems will be reported elsewhere. A particularly interesting application would be to the two-dimensional Hubbard model away from half-filling where it has been shown using a type of static approximation that at $T$ = 0 long-range order makes way to phases with short-range magnetic order only~\cite{Trapper1995}. It would be very important to know whether these results survive if the spin dynamics is taken into account using the dyn-NLCPA.

%--------------------------------------------------------------------------------------------------------------------------------------------
\ack

D.A.R. would like to thank Prof.~Y.~Kakehashi for discussions and for his hospitality while visiting the University of the Ryukyus, Okinawa. This work is supported by National Natural Science Foundation of China (Nos. 11174219 and 11474217), Program for New Century Excellent Talents in University (NCET-13-0428), Research Fund for the Doctoral Program of Higher Education of China (No. 20110072110044) and the Program for Professor of Special Appointment (Eastern Scholar) at Shanghai Institutions of Higher Learning as well as the Scientific Research Foundation for the Returned Overseas Chinese Scholars, State Education Ministry.

%--------------------------------------------------------------------------------------------------------------------------------------------
\section*{References}

%\bibliographystyle{prsty}
%\bibliography{nldcpa}

%--------------------------------------------------------------------------------------------------------------------------------------------
\end{document}